\documentclass{pasj01}
\Received{$\langle$2021 November 18$\rangle$}
\Accepted{$\langle$2021 December 22$\rangle$}
\usepackage{ulem}
\usepackage{lineno}

\begin{document}

\title{Discovery of Stable Titanium at the Northeastern Jet of Cassiopeia A: Need for a Weak Jet Mechanism?}

\author{Takuma Ikeda\altaffilmark{1}, Yasunobu Uchiyama\altaffilmark{2,1}, Toshiki Sato\altaffilmark{1,*}, Ryota Higurashi\altaffilmark{1}, Tomoya Tsuchioka\altaffilmark{1}, Shinya Yamada\altaffilmark{1}}

\altaffiltext{1}{Department of Physics, Rikkyo University,  3-34-1 Nishi Ikebukuro, Toshima-ku, Tokyo 171-8501, Japan}
\altaffiltext{2}{Graduate School of Artificial Intelligence and Science, Rikkyo University,  3-34-1 Nishi Ikebukuro, Toshima-ku, Tokyo 171-8501, Japan}
\maketitle
\email{toshiki.sato@rikkyo.ac.jp}


\maketitle

\begin{abstract}

The origin of the jet-like structures observed in Cassiopeia A is still unclear, although it seems to be related to its explosion mechanism. X-ray observations of the characteristic structures could provide us useful information on the explosive nucleosynthesis via the observation of elements, which is a unique approach to understand its origin. We here report the discovery of shocked stable Ti, which is produced only at the inner region of exploding stars, in the northeast jet of Cassiopeia A using the 1-Ms deep observation with the {\it Chandra} X-ray observatory. The observed Ti coexists with other intermediate-mass elements (e.g. Si, S, Ar, Ca) and Fe at the tip of the X-ray jet structure. We found that its elemental composition is well explained with the production by the incomplete Si burning regime, indicating that the formation process of the jet structure was sub-energetic at the explosion (the peak temperature during the nuclear burning must be $\lesssim$ 5$\times$10$^{9}$ K at most). Thus, we conclude that the energy source that 
formed the jet structure was not the primary engine for the supernova explosion. Our results are useful to limit the power of the jet-structure formation process, and a weak jet mechanism with low temperature may be needed to explain it.

\end{abstract}
%\linenumbers

\section{Introduction}
Asymmetry is widely believed to play an important role in the explosion mechanisms of core-collapse supernovae (CC SNe).
Recent simulations of the neutrino-driven explosion 
that is a currently viable mechanism \citep{1985ApJ...295...14B} have indicated that the SN shock wave can be accelerated mainly by convective hot bubbles produced by neutrino heating and standing accretion shock instability: SASI \citep{2012ARNPS..62..407J,2021Natur.589...29B}. 
These effects could be related to the observed asymmetric ejecta distributions and neutron-star kick directions of CC remnants \citep{2011ApJ...732..114L,2017ApJ...844...84H,2020ApJ...889..144H,2018ApJ...856...18K}. In addition to this mechanism, bipolar explosions that could originate from the rotation of the progenitor star are also believed to be an important asymmetric mechanism to explain SN/SNR observations \citep{1999ApJ...524L.107K,2000ApJ...541.1033F,2009ApJ...691.1360T}. In this case, extremely bimodal asymmetry on the ejecta distribution is expected, which is a major feature of this mechanism. To unveil the SN explosion mechanisms, it would be useful to test these different asymmetric effects using observations.

The Galactic SNR Cassiopeia A (Cas A) provides us a great opportunity to probe the asymmetric effects at the explosion. For example, the high-velocity optical knots have clearly shown the asymmetric explosion in Cas A \citep{1996ApJ...470..967F,2001ApJS..133..161F,2006ApJ...645..283F,2016ApJ...818...17F,2001AJ....122..297T}. In particular, the S-rich optical knots in the NE jet and SW counterjet regions have high velocities of $>$ 10,000 km s$^{-1}$, which is a unique feature connected to the explosion mechanism. These jet-like structures are kinematically and chemically distinct from the rest of the remnant, which seems to be related to the explosion mechanism. In addition, the bubble-like interior observed in Cas A also indicates strong evidence of asymmetric mixing processes during the explosion \citep{2015Sci...347..526M}. Not only these characteristic asymmetric structures, the observations of some specific elements such as $^{44}$Ti, stable Ti and Cr synthesized in the high-entropy nuclear burning regime also allow us to test asymmetric activities at the heart of the SN explosion of Cas A \citep{2014Natur.506..339G,2017ApJ...834...19G,2021Natur.592..537S}. As above, the asymmetric explosion of the Cas A supernova has been well discussed from various viewpoints and its origin has been debated for years.

Recent theoretical and observational studies have supported that the asymmetries by the neutrino-driven mechanism can explain well the observational properties of Cas A \citep{2000ApJ...528L.109H,2014Natur.506..339G,2017ApJ...834...19G,2017ApJ...842...13W,2021A&A...645A..66O,2020ApJ...895...82V,2021Natur.592..537S}. On the other hand, the characteristic jet-like structures cannot be explained by only this mechanism, and its formation process is still an open question. Here, we would note that the jet-like structures do not appear to have been created by a bipolar explosion (in other words, the structures did not drive the explosion). In bipolar explosions, Fe-rich materials that are produced by $\alpha$-rich freezeout should be ejected at high velocities along the jet axis \citep{2003ApJ...598.1163M,2009ApJ...690..526T}, however the existence of such Fe-rich ejecta have not been reported in this region \citep{Hwang2004}. One of candidates for the jet formation process would be some post-explosion activities. Some previous studies have suggested that an underenergetic jet from a rotating new-born neutron star, which would have emerged after the supernova explosion, might explain this feature \citep{2005ASPC..332..350B,2021A&A...645A..66O}. However, there is not firm observational evidence on it.

In order to understand its origin, we investigate the elemental composition in the northeastern jet structure in this paper. If the structure was produced in the Si-burning regime around the supernova engine (in other words, if it was related to the inner asymmetries), we can discuss how deep into the supernova it was produced, by measuring its elemental composition \citep{1973ApJS...26..231W,2017ApJ...834..124Y,2020ApJ...890..104S,2021Natur.592..537S,2021arXiv210504101O}. For example, the elemental composition enriched in the intermediate mass elements (e.g. Si, S, Ar, Ca) and Fe have already been suggested to be of Si-burning origin \citep{Hwang2004}. However, a more detailed investigation of the elements and comparison with nucleosynthesis calculations are still necessary to clarify its origin. In this research, we focus on the search for Ti and other Fe-group elements that can characterize that burning layer and discuss the origin of the jet structure.

%%%%%%%%%%%%%%%%%%%%%%%%%%%%%%%%%%%%%%%%%%%%%%%%%%%%%%%%%%%%%%%%
%%%%%%%%%%%%%%%%%%%%%%%%%%%%%%%%%%%%%%%%%%%%%%%%%%%%%%%%%%%%%%%%

\begin{figure}[t]
\begin{center}
\includegraphics[bb=0 0 1356 1162, width=0.485\textwidth]{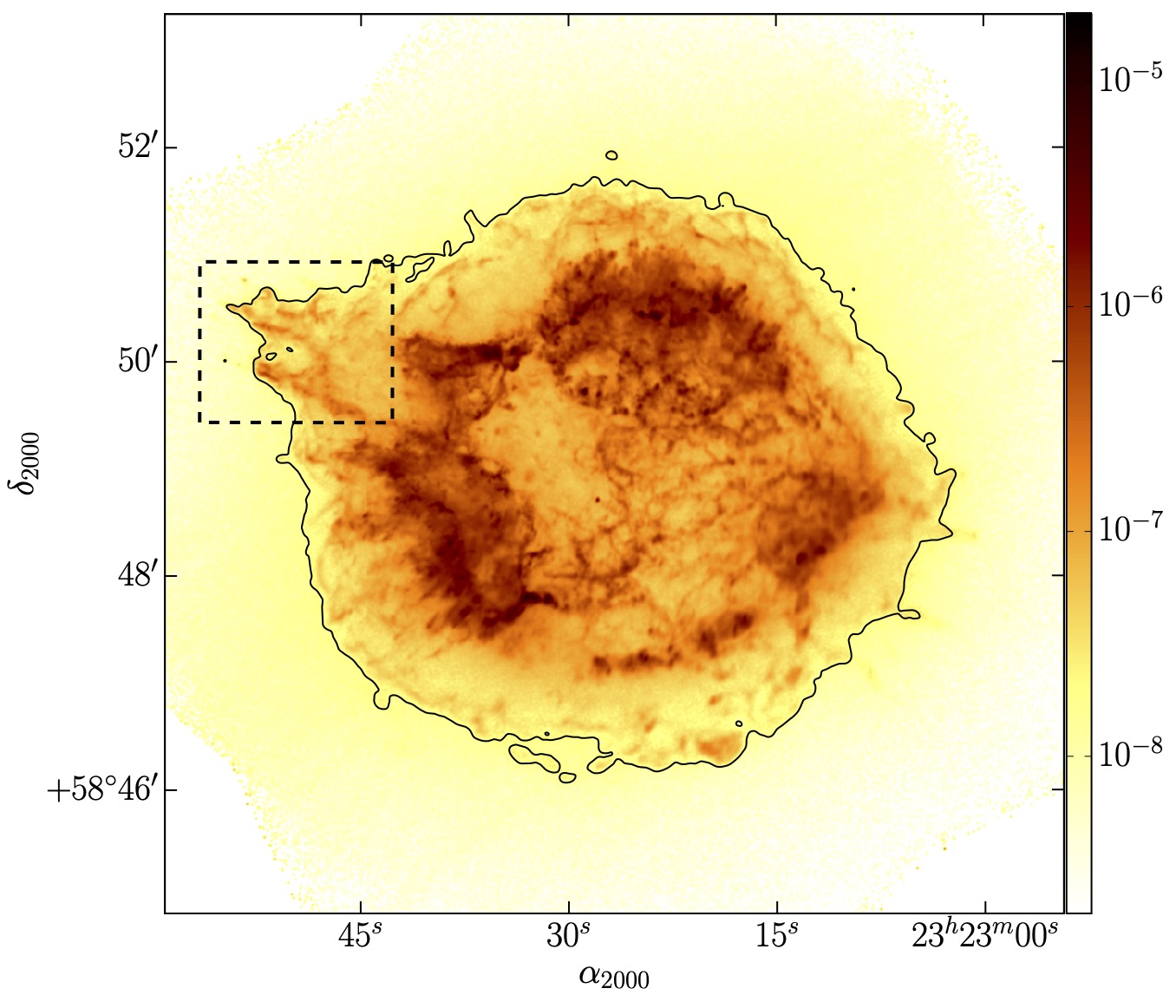}
\end{center}
\caption{The {\it Chandra} X-ray image of Cassiopeia A in the 1.75-1.95 keV band smoothed with a Gaussian function with a sigma of 1 pixel (0.492$^{\prime\prime}$). The edge of the remnant is traced with the black curve. The black broken rectangle shows the northeastern jet region that is the focus of this study.}
\label{Si_map}
\end{figure}

\begin{figure*}[t]
\begin{center}
\includegraphics[bb=0 0 2160 1074, width=1.\textwidth]{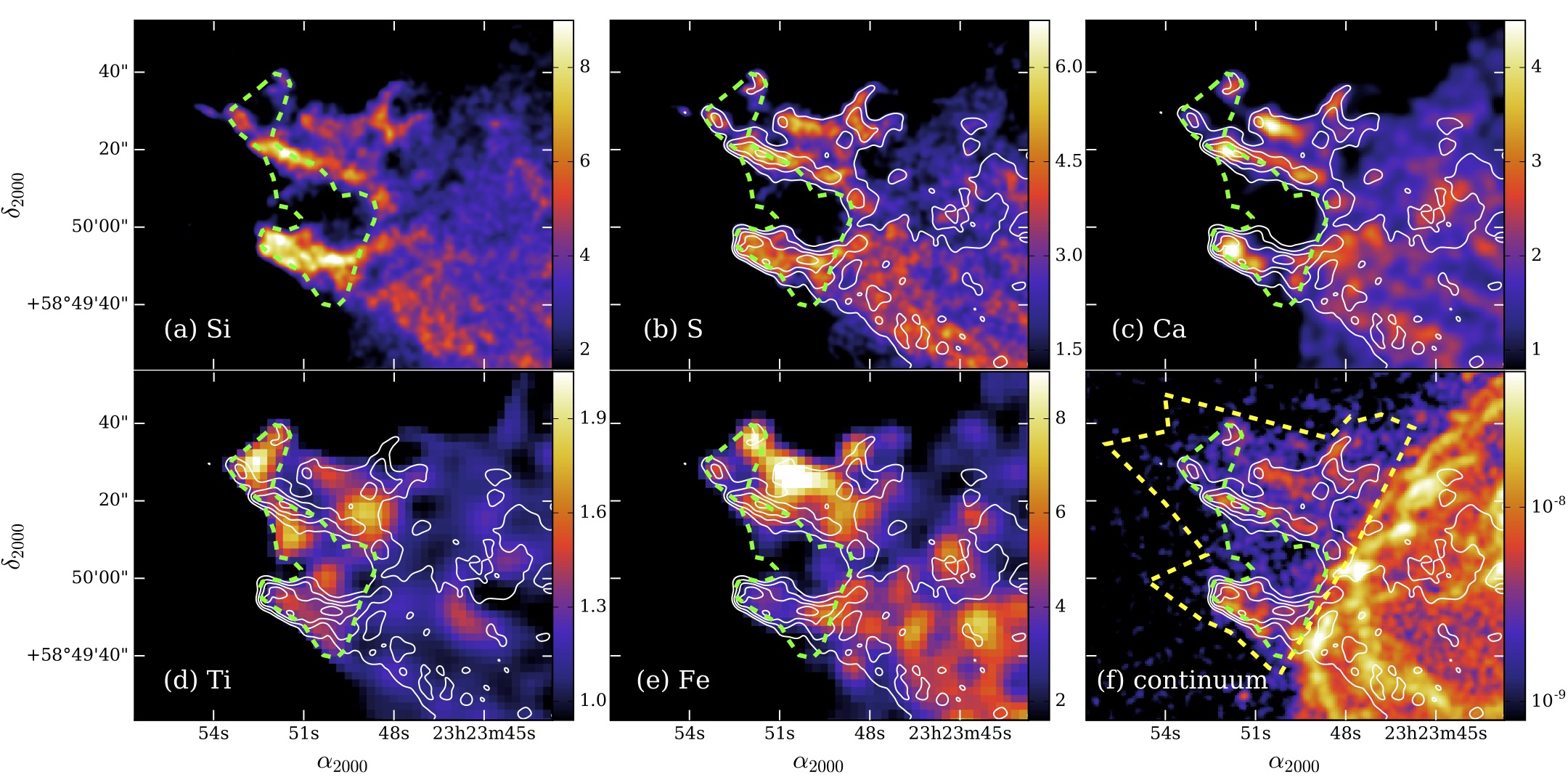}
\end{center}
\caption{Elemental distributions around the NE jet region of Cassiopeia A. Each line image is normalized by a image of continuum emissions taken from nearby energy bands. (a) Si map obtained in the 1.75-1.95 keV band. We superimposed the regions selected for the spectral analysis (in green and yellow) together with the contour levels of the Si line ratio image at 20\%, 30\%, 40\%, and 50\% of the maximum (in white) for (b)-(f) panels. (b) S map obtained in the 2.35-2.55 keV band. (c) Ca map obtained in the 3.75-4.00 keV band. (d) the Ti map obtained in the 4.65-4.82 keV band. (e) Fe line ratio map obtained in the 6.48-6.80 keV band. (e) count-rate images in the 3.45-3.55, 4.2-4.4, 5.1-5.5, 5.8-6.0 keV. The bin sizes of the images are 0.246$''$ for (a), (b), 0.492$''$ for (c), (f), and 1.967$''$ for (d), (e), respectively.}
\label{element_map}
\end{figure*}

\section{Observation and Image Analysis}
\subsection{Observation and Data Reduction}
    Cas A has been observed multiple times with {\it Chandra} X-ray observatory since it was observed as first light target of {\it Chandra} in 2000 \citep{2000ApJ...537L.119H,Hwang2004,2000ApJ...528L.109H,2008ApJ...677L.105U,2011ApJ...729L..28P,2014ApJ...789..138P,2017ApJ...836..225S,2018ApJ...853...46S}. The ejecta positions are shifting from observation to observation due to its expansion. Therefore, we used only the deepest observation in 2004 composed of separate 9 ObsIDs (4634, 4635, 4636, 4637, 4638, 4639, 5196, 5319, 5320) \citep{Hwang2004} to reduce the uncertainty on the region selection. The data were reprocessed using CIAO 4.8 and {\it Chandra} CalDB version 4.7.0 In general, the data quality depends on the condition of individual observations because data sometimes suffer from particle flares in orbit which can not be removed by the standard pipeline process. These particle flares cause high background in the detector, we exclude such intervals by using light curve in off-source for each observation. Time periods where the count rate deviates from the mean value during quiescent periods by $\geq 3 \sigma$ are removed from subsequent analysis. These filtered data which range between $\sim$40 ks and $\sim$160 ks and total exposure of them are $\sim$960 ks. 
    All the spectra were extracted individually for each of the nine observation segments, and corresponding detector response files were generated. The individual spectra were then added together to produce the final spectrum for a particular region, while the individual response files were weighted according to the relative exposure time of each observation segment before being added to produce the final response files.

%%%%%%%%%%%%%%%%%%%%%%%%%%%%%%%%%%%%%%%%%%%%%%%%%%%%%%%%%%%%%%%%
%%%%%%%%%%%%%%%%%%%%%%%%%%%%%%%%%%%%%%%%%%%%%%%%%%%%%%%%%%%%%%%%

\subsection{Distribution of Elements in the Northeast Jet}
     Cas A is observed as a complex mixture of thermal and non-thermal X-ray radiations. Thus, we would need to perform detailed image analysis to extract pure ejecta regions associated with the jet from the observed data.
     The northeastern (NE) jet-like structure of Cas A can be observed as a Si-rich feature in X-rays (Figure \ref{Si_map}), which is the brightest jet-like structure in the remnant. Here we can see the Si-rich filaments in the northeast region beyond the forward shock clearly. Although the counter jet structure in the southwest has been also known, this study focuses only on the NE jet. This is because that the strong non-thermal radiation and weak thermal radiation in this region. \par
     
     In order to clarify more detailed spatial distribution of elements in the NE jet region, we produced ratio maps between K-shell line emissions and continuum emissions for each element (Figure \ref{element_map}). As a result, we found that the intermediate mass elements (Si, S, Ca, Ti) and Fe seem to coexist along the jet structure, showing clumpy/filamentary distribution. Although the distribution of elements and the plasma parameters around the jet region has been investigated well \citep{Hwang2004,Laming2006}, our new maps provide us the most detailed ones. In addition, we would emphasize that the Ti map is for the first time produced in this research. Strictly speaking, there also exists emission lines from Ca in the energy band for the Ti map (4.65--4.82 keV). On the other hand, the difference from the Ca map implies a possibility that they have different origins, supporting a real distribution of the shocked Ti. To discuss its elemental composition accurately, we have performed spectral analysis in following sections. 
     \par

\section{Spectral Analysis}
\subsection{Detection of Ti–K$\alpha$ emissions}

    In the previous section, we found the Ti emissions seem to be concentrated at the tips of the NE jet region. In theory, Ti should be produced at the inner area of exploding stars \citep{Thielemann1996,2001ApJ...555..880N}. Therefore, the jet may be related to the central activity at the explosion. On the other hand, we would need a more careful investigation using spectroscopy to verify the existence of Ti. Here, we show detail investigations of the Ti emissions using spectral analysis.
    
\begin{figure}[t!]
\begin{center}
\includegraphics[bb=0 0 1046 1020, width=0.46\textwidth]{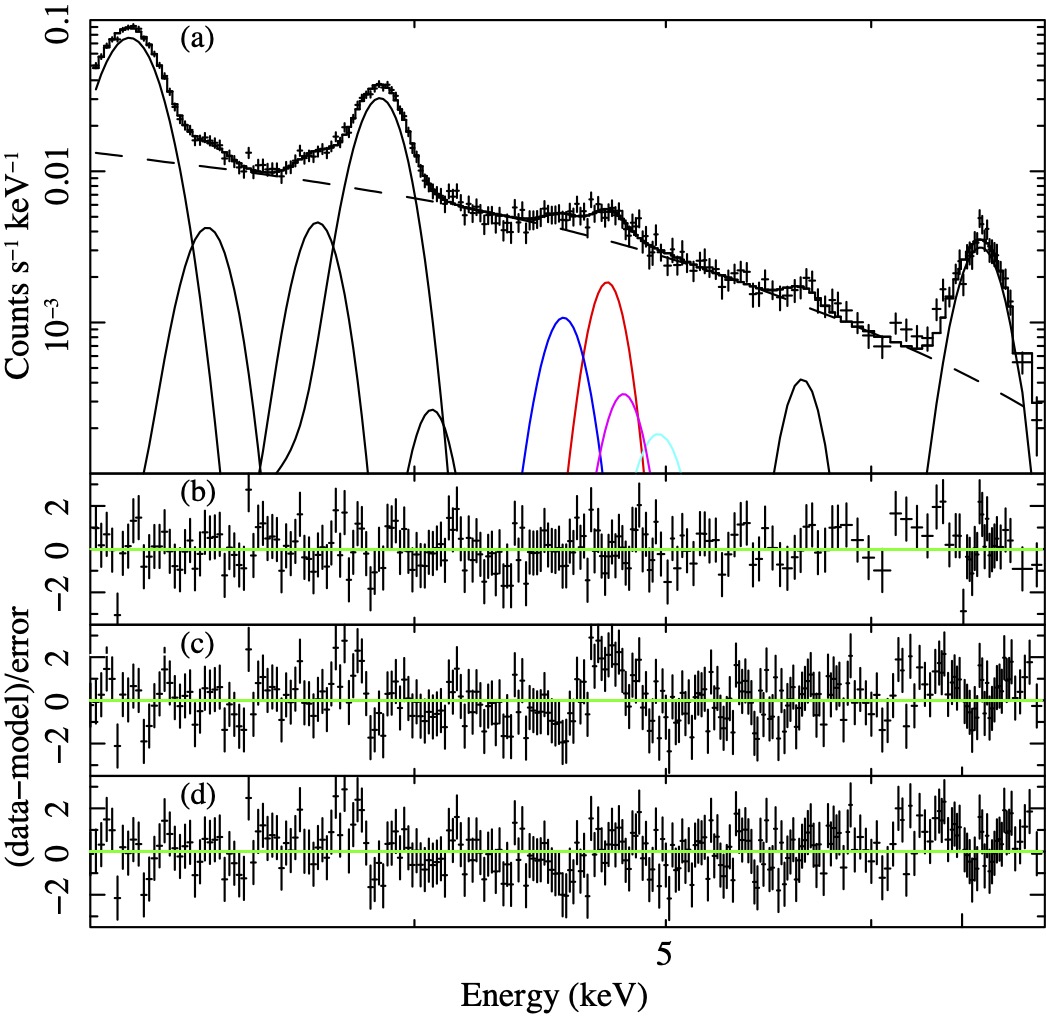}
\end{center}
\caption{(a) ACIS spectrum in the 3-7 keV band from the green region of Figure \ref{element_map}. The best-fit bremss model (dot line) and lines of Ar, Ca, Ti, Cr, and Fe with Gaussian functions are shown by solid lines, but Ca He$\beta$ (blue), Ti K$\alpha$ (red), Ca He$\gamma$ (magenta), and Ca He$\delta$ (cyan). (b) The residual from the best-fit model of (a). (c)  Residual from the best-fit vvpshock model with solar Ti abundance. (d) Residual from the same model as (c), except that the Ti abundance is taken into account. }
\label{spectrum_ti}
\end{figure}

    Figure \ref{spectrum_ti} shows the ACIS-S spectrum extracted from the NE jet region (green region in Fig.\ref{element_map}). As a result, we have found a strong Ti-K$\alpha$ line in the spectrum at a confidence level greater than 5$\sigma$. We here used two kinds of models to fit the spectrum: (1) an absorbed single-component plane-parallel shock model ({\it vvpshock} in XSPEC) and (2) a thermal bremsstrahlung emission ({\it bremss} in XSPEC) with 10 Gaussian models. The best-fit parameters are summarized in Table \ref{tab:jet_y_ti_h}. We found the quality of the fits improves greatly in both models by adding Ti component ($\Delta \chi^2$ = 34, 40 in the vvpshock fit and the bremss-Gausssian fit, respectively), indicating that the Ti-K detection is 5.3, 5.6$\sigma$ level. Interestingly, the Ti abundance is significantly higher than the solar value, which would be useful to discuss the origin of the jet structure (the detailed discussion is in the Discussion section). \par 
   
   In the energy band around the Ti emissions, there also exists some weak Ca lines (i.e. Ca K$\beta$ and K$\gamma$). We have carefully evaluated these emissions in all our models and have concluded that the handling of these weak lines does not change our conclusion. For example, the Doppler effect of these lines up to 10-20 eV may explain 4.7 keV feature in the spectrum. However, with such a large shift, the Ca He-like line should be also wider (or distorted) than observed. In addition, the observed Ti line is too strong to be explained with the other components. In fact, there does not exist any strong lines except Ti-K$\alpha$ emissions at just between the Ca K$\beta$ and K$\gamma$ emissions (see also Extended Data Fig.2b in \cite{2021Natur.592..537S}).
   
   We have also examined influence of pile-up events on the spectrum. Actually, the pile-up effect can not ignore at some bright spots in Cas A. In particular, strong line emissions such as Si He$\alpha$ and S He$\alpha$ make fake line features at specific energies (e.g. Si~He$\alpha$ + Si~He$\alpha \approx $ 3.7 keV, Si~He$\alpha$ + S~He$\alpha \approx$ 4.3 keV). However, the 4.7 keV emissions exist only at the NE jet region where the Si and S emissions are not so strong. In addition, there does not exist any combination of lines to explain the 4.7 keV emissions. Moreover, the line centroid energy of 4.7 keV corresponds to the emissions from the He-like Ti ions, which means the ionization state of the shocked Ti agrees well with the other elements. From all these facts, we conclude that the 4.7 keV emissions originate from the shocked Ti and our measurement is robust.

\begin{table}[t]
\tbl{Spectral parameters of models for jet spectrum of Cassiopeia A.}{
\begin{tabular}[t]{llll}\hline\hline
\multicolumn{4}{c}{BREMSS+Gausssian} \\%\cline{1-4}%\hline
\multicolumn{4}{l}{Continuum} \\
\multicolumn{2}{l}{kT [keV]} & \multicolumn{2}{l}{2.40$^{+0.24}_{-0.23}$} \\
\multicolumn{2}{l}{Normalization\footnotemark[a]} & \multicolumn{2}{l}{5.68$^{+1.21}_{-0.81}$}\\ \hline
Line\footnotemark[b]& E$\rm _{Center}$&Line Width & Flux \\
& (keV)& ($10^{-2}$ keV)& ($10^{-6}$ cm$^{-2}$ s$^{-1}$) \\
ArHe$\alpha$& 3.11$\pm 0.00$&4.05$\pm 0.37$&42.1$^{+1.2}_{-1.4}$ \\
ArLy$\alpha$& 3.34$\pm 0.02$&0 (fixed)&1.87$^{+0.40}_{-0.48}$ \\
ArHe$\beta$& 3.67$^{+0.03}_{-0.01}$&0 (fixed)&1.96$^{+0.35}_{-0.37}$ \\
CaHe$\alpha$& 3.88$\pm 0.00$&3.64$^{+0.57}_{-0.83}$&15.5$^{+0.5}_{-0.8}$ \\
CaLy$\alpha$\footnotemark[c]& 4.06 (fixed)&0 (fixed)&0.111$^{+0.270}$ \\
CaHe$\beta$\footnotemark[c]& 4.57 (fixed)&3.57 (fixed)&0.596$^{+0.209}_{-0.251}$ \\ 
Ti K$\alpha$& 4.75$^{+0.02}_{-0.04}$&0 (fixed)&0.937$^{+0.274}_{-0.263}$ \\
Ca He$\gamma$\footnotemark[c]&4.82 (fixed)&3.36 (fixed)&0.204$^{+0.236}$ \footnotemark[d] \\
Ca He$\delta$\footnotemark[c]&4.97 (fixed)&6.17 (fixed)&0.143 \\
Cr K$\alpha$\footnotemark[c]&5.64 (fixed)&4.30 (fixed)&0.397$^{+0.187}_{-0.210}$ \\
Fe K$\alpha$&6.61$\pm 0.01$&8.84$^{+1.19}_{-1.35}$&7.17$\pm 0.50$  \\ 
\multicolumn{1}{l}{$\chi ^2$/d.o.f}&&\multicolumn{2}{l}{273.6/228 $\longrightarrow$ 233.9/226} \\ \hline\hline
\multicolumn{4}{c}{NEI (VVPSHOCK)} \\ 
&&Solar Ti&Free Ti\\
kT [keV]&&2.36$^{+0.19}_{-0.22}$&2.26$^{+0.17}_{-0.20}$\\
Tau&&1.75$^{+0.19}_{-0.18}$&1.89$^{+0.17}_{-0.20}$\\
\multicolumn{2}{l}{(10$^{11}$ s cm$^{-3}$)}&&\\
\multicolumn{2}{l}{Normalization\footnotemark[a]}&8.03$^{+0.52}_{-0.54}$&7.70$\pm 0.56$\\
Z(Ar/Ca)&&0.81$^{+0.09}_{-0.12}$&0.83$^{+0.11}_{-0.13}$ \\
Z(Ti/Ca)&&&3.78$^{+1.23}_{-1.48}$ \\
Z(Cr/Ca)&&0.82$^{+0.49}_{-0.59}$&0.94$^{+0.52}_{-0.63}$ \\
Z(Fe/Ca)&&0.29$^{+0.05}_{-0.07}$&0.30$^{+0.06}_{-0.10}$ \\
Z$_{\rm Ca}$&&18.7$^{+1.44}_{-1.87}$&19.9$^{+1.75}_{-2.19}$ \\
\multicolumn{1}{l}{$\chi ^2$/d.o.f}&&\multicolumn{2}{l}{300.5/239 $\longrightarrow$ 266.8/238} \\ \hline
\end{tabular}}\label{tab:jet_y_ti_h}
\begin{tabnote}
\par \hangindent3pt \noindent
\hbox to24pt{{\bf NOTE.} \hss}\unskip%
The errors given here mean 90\% confidence limits.

\par \noindent
\hbox to6pt{\footnotemark[a]\hss}\unskip%
The unit is 10$^{-18}$ $\int n_en_H$dV /(4$\pi D^2$) (cm$^{-5}$), where $n_e$, $n_H$ V, and $D$ are the electron and hydrogen densities (cm$^{-3}$), emitting volume (cm$^{3}$), and distance to the source (cm), respectively.

\par \noindent
\hbox to6pt{\footnotemark[b]\hss}\unskip%
Considering redshift z = $-$1.94$\times 10^{-3}$ for all Gaussians.

\par \noindent
\par \hangindent6pt \noindent
\hbox to6pt{\footnotemark[c]\hss}\unskip%
Using plasma code value for line's energy and width.

\par \noindent
\hbox to6pt{\footnotemark[d]\hss}\unskip%
Lower error is unconstrained.
\end{tabnote}
\end{table}

\begin{figure}[t!]
\begin{center}
\includegraphics[bb=0 0 1096 766, width=0.45\textwidth]{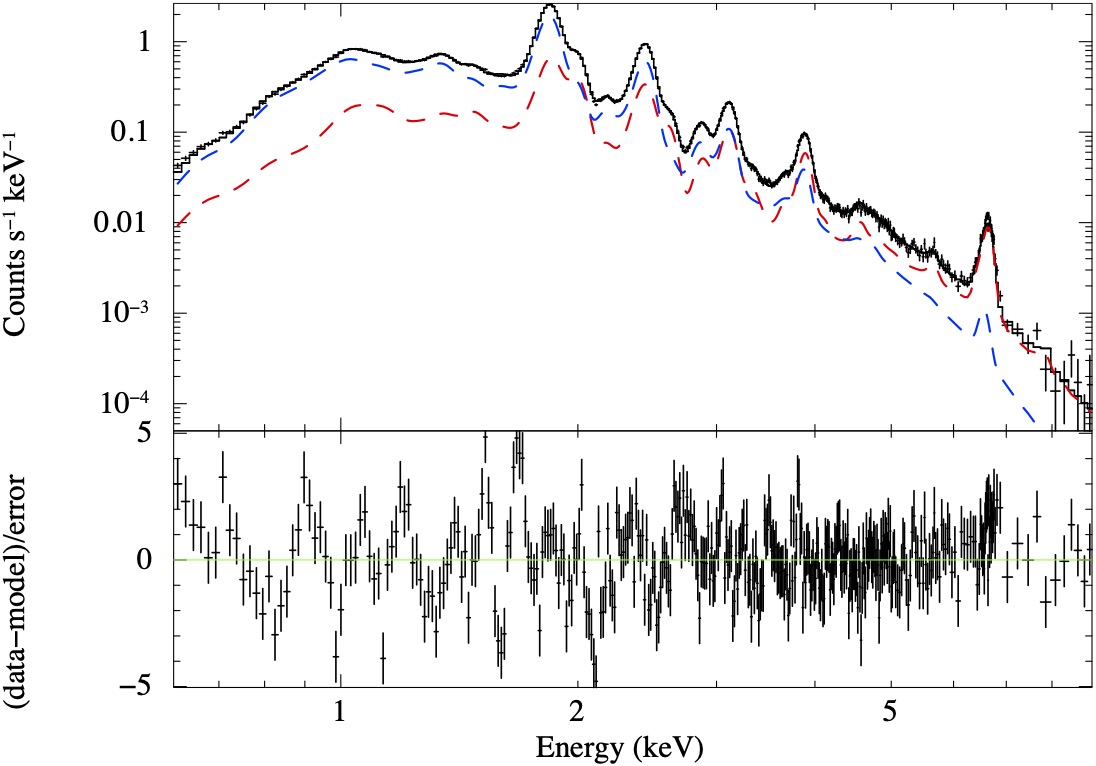}
\end{center}
\caption{The X-ray spectrum and the best-fit model (black solid) in the entire NE jet region using the two-temperature model. The red and blue broke lines show the hot and cool vvpshock models, respectively. The N abundance is fixed to 100 in this model.}
\label{two_temp_model}
\end{figure}

\subsection{Titanium Mass in the NE Jet}
     The total mass of the observed Ti would also provide us important information for discussing where the jet structure was formed during the explosion. In order to estimate the mass of the observed Ti in the NE jet, we here fit the X-ray spectrum with two non-equilibrium ionization plasma models ({\it vvpshock} model in XSPEC; \cite{Borkowski2001}) assuming the interstellar absorption ({\it wabs} in XSPEC). In this analysis, we used the entire jet region just outside forward shock (yellow region of Figure \ref{element_map}).
    
    We summarized the best-fit abundances obtained from our new fitting in Table \ref{tab:jet_abun}. Our model explained the spectrum well (Fig.\ref{two_temp_model}; reduced $\chi ^2 <$ 2) and we found the plasma parameters of $kT_{\rm cool}$ $\sim$ 1.3 keV, $kT_{\rm hot}$ $\sim$ 9 keV, $n_{\rm e}t_{\rm cool}$ $\sim$ 1.8$\times 10^{11}$ s cm$^{-3}$, $n_{\rm e}t_{\rm hot}$ $\sim$ 1.1$ \times 10^{11}$ s cm$^{-3}$. We would note that the estimation of the electron density and mass depends on the amount of lighter elements that are difficult to measure.  Therefore, we performed some spectral fits assuming different abundances for such a light element in this analysis. We here assumed a N-rich plasma for both the hot and cool components because the optical knots dominated by [N II] along the NE jet were reported from the HST observations \citep{Hammell2008}. We fitted the spectrum between 0.6 and 9 keV with varying the N abundance from 100 to 12800 (solar ratio), and estimated its electron density and Ti mass. The abundances of O, Ne, Na, Mg, Al, Si, S, Ar, Ca, Ti, Cr, Fe and Ni were tied between the hot and cool components and fitted as free parameters, while the rest of the elemental abundances were fixed at the solar values \citep{An_Gre}. In this way, we estimated upper and lower bounds (90\% level) of electron density and Ti mass. \par
    
    The spectral fitting provides the electron temperature $kT_e$, emission integral $E_I = \int n_en_HdV$, ionization age $I = \int n_edt$ and elemental abundances Z for two plasma components in the jet of Cas A, where $n_e$ and $n_H$ are the densities of electron and hydrogen, respectively. For measuring Ti mass, we calculate the number of electrons per hydrogen atom $R_e = n_e/n_H$ and the effective number of protons and neutrons (baryon mass) per hydrogen atom $R_m = n_m/n_H$ from the fitting parameters (see also  \cite{Willingale2003}), where $n_m$ is total number density of protons+neutrons (including those bound up in nuclei). Here we assume a fully ionised plasma. Thereafter, we can estimate the electron density, hydrogen density, Ti mass ($M_{\rm Ti}$), and thermal pressure as following:
%    \begin{fleqn}[0pt]
    \begin{eqnarray}
    %\begin{align}
    &&n_e = \sqrt{E_IR_e/(V\eta)} \\
    &&n_H = \sqrt{E_I/(V\eta R_e)} \\
    &&M_{\rm Ti} = m_{\rm Ti}Z_{\rm Ti}\sqrt{E_I\eta V/R_e} \\
    &&P_{th} = k(T_iR_i + T_eR_e)\sqrt{E_I/(V\eta R_e)}
    %&M_{H}/M = 1/(\Sigma Z_{all}m_z)
    \end{eqnarray}
    %\end{align}
%    \end{fleqn}
    Here, $m_{\rm Ti}$ is the mass of Ti atom, $V$ is the total plasma volume, $\eta$ is a filling factor within that volume and $Z_{Ti} = n_{Ti}/n_H$ is the abundance of Ti. Assuming the distance of 3.4 kpc \citep{Reed1995} and a cylinder volume with a 1$^{\prime\prime}$ diameter for expressing the jet structure, the volume of the NE jet can be estimated to be $\sim$1.5$\times 10^{55}$ cm$^3$ from the projected jet area ($\sim$0.65 arcmin$^2$). 
    To calculate the mass associated with the hot and cool components, we must estimate the filling factors. We here assumed pressure equilibrium between the hot and cool plasma, satisfying $(T_iR_i/T_eR_e)_{\rm hot} = (T_iR_i/T_eR_e)_{\rm cool}$, to give an estimate of the filling factors within the overall jet. As we noted, the value of electron density and mass have a large uncertainty from fully ionized light nuclei. \citet{Laming2006} reported some of the jet stem knots fall at lower temperatures than modeled assuming O to be the most abundant element, which could be understood the jet knot plasma comprises a substantial fraction of H. Therefore, we expected hydrogen occupy half of the mass of the jet plasma ($1/R_m <0.5$) as reasonable limits. Taking all this into account, the filling factor for the components $\eta _{hot}$ and $\eta _{cool}$ are assumed to be 0.94 and 0.06, respectively. As a result, the Ti mass for the hot and cool component has been estimated to be in the range of (0.30--1.7)$\times 10^{-5}$ $M_{\solar}$ and (0.14--0.73)$\times 10^{-5}$ $M_{\solar}$ with 90\% confidence range, where the electron densities of each component are 4.5--15 cm$^{-3}$ and 0.6--2.2 cm$^{-3}$, respectively. Eventually, we obtained the total Ti mass is  (0.44--2.4)$\times 10^{-5}$ $M_{\solar}$ as the sum of the hot and cool components.
    
\begin{table}[t]
\tbl{Results of the spectral analysis for the whole of Jet.}{
\begin{tabular}{lll} \hline \hline
%\begin{tabularx}{lccc||llcc} \hline \hline
 & Solar Abundance\footnotemark[a] & Mass [$M_\odot$]\\ \hline
O/Ca & 0.37 $\pm$ 0.06 & 56 $\pm$ 9 \\
Ne/Ca & 0.063 $\pm$ 0.010 & 1.7 $\pm$ 0.3 \\
Na/Ca & 1.6 $\pm$ 0.2 & 0.84 $\pm$ 0.11 \\
Mg/Ca & 0.065 $\pm$ 0.007 & 0.65 $\pm$ 0.07 \\
Al/Ca & 0.19 $\pm$ 0.02 & 0.71 $\pm$ 0.08 \\
Si/Ca & 0.55 $\pm$ 0.08 & 6.0 $\pm$ 0.8 \\
S/Ca & 0.72 $\pm$ 0.10 & 4.1 $\pm$ 0.6 \\
Ar/Ca & 0.73 $\pm$ 0.09 & 1.0 $\pm$ 0.1 \\
Ti/Ca & 1.4 $\pm$ 0.4 & 0.073 $\pm$ 0.020 \\
Cr/Ca & 0.51 $\pm$ 0.14 & 0.13 $\pm$ 0.04 \\
Fe/Ca & 0.13 $\pm$ 0.02 & 3.7 $\pm$ 0.6 \\
Ni/Ca & 0.12 $\pm$ 0.046 & 0.13 $\pm$ 0.05 \\
 \hline
\end{tabular}}\label{tab:jet_abun}
\begin{tabnote}
\par \hangindent3pt \noindent
\hbox to24pt{{\bf NOTE.} \hss}\unskip%
The errors mean 90\% confidence limits.
\end{tabnote}
\end{table}

\begin{figure*}[t!]
\begin{center}
\includegraphics[bb=0 0 1554 580, width=0.96\textwidth]{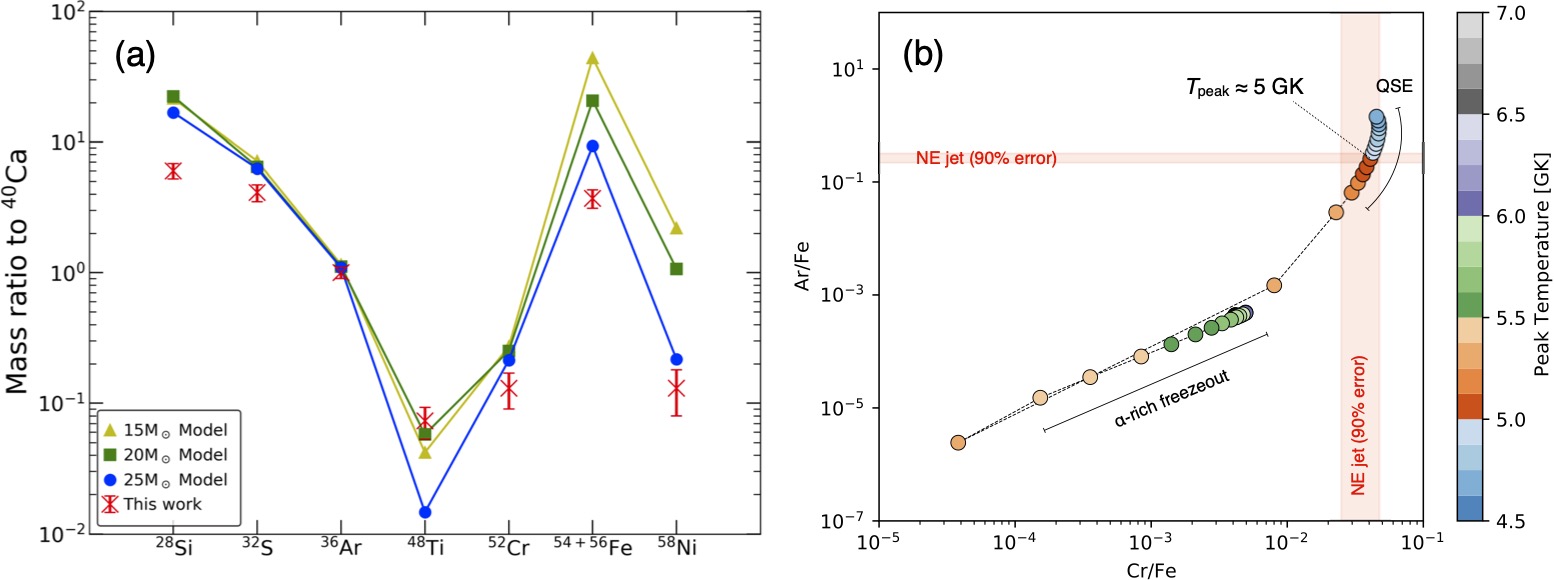}
\end{center}
\caption{(a) Observational result and theoretical calculations of element mass ratio to calcium. The observed values are same as shown in Table \ref{tab:jet_abun}, and compared to spherical explosive nucleosynthesis model in \citet{Thielemann1996} of progenitor mass of 15 M$_{\solar}$, 20 M$_{\solar}$, and 25 M$_{\solar}$ with Y$_e$ is constant. Note that isotope of element indicate just model result while it is generally uncertain from observation. (b) The observed Cr/Fe and Ar/Fe mass ratios compared with a 1D SN model in \citet{2020ApJ...893...49S}.}
\label{el_ca}
\end{figure*}

\section{Discussion}

    In this research, we for the first time have detected the Ti-K line from the NE jet structure of Cas A and estimated the total Ti mass of (0.44--2.4)$\times 10^{-5}$ $M_{\solar}$. The detection of the rare element that could have been produced by the high-entropy burning process is timely \citep{2014Natur.506..339G,2021Natur.592..537S}, and this new observation would help us to understand the origin of the Cas A jet. In this section, we discuss the origin of Ti in the jet region and how it is related to the explosion mechanism.
    
    \subsection{The Origin of the Titanium Emissions}
    If the observed Ti was surviving $^{44}$Ti before its decay, the jet formation could be related to some strong central activity during the explosion \citep{1998ApJ...492L..45N,2003ApJ...598.1163M}. However, with just only the detection of the Ti line with X-rays, we cannot conclude whether the observed Ti is radioactive $^{44}$Ti or stable ones, since X-ray observations cannot distinguish among different isotopes. 
    Here, the observations of $^{44}$Ti from Cas A \citep{1994A&A...284L...1I,2001ApJ...560L..79V,2006ApJ...647L..41R,2014Natur.506..339G} provide us useful information to discuss it.
    The total amount of $^{44}$Ti in Cas A has been estimated to be (1.25$\pm$0.3$)\times$10$^{-4}$ $M_{\solar}$ from the {\it NuSTAR} observation \citep{2014Natur.506..339G}. Assuming the remnant age of 340 yrs and the half life of 60 yrs for $^{44}$Ti, the total $^{44}$Ti mass at the present time is estimated to be $\approx$2.5$\times$10$^{-6}$ $M_{\solar}$, which is much smaller than the total mass of the shocked Ti we observed.
    Furthermore, at least 80\% of the observed $^{44}$Ti emission is contained within the reverse shock radius as projected on the plane of the sky \citep{2014Natur.506..339G}. Thus, the estimated Ti mass in the NE jet (outside of reverse shock) is roughly an order of magnitude larger than the expectation from the $^{44}$Ti observation, suggesting the $^{44}$Ti is not the main source for the observed Ti.
    Also, we estimated the 68 keV line flux from $^{44}$Ti in the NE region of Cas A using the {\it NuSTAR} archive data and set the upper limit of $^{44}$Ti mass of $\approx10^{-7}$ M$_{\solar}$ in the NE jet, which is much lower than that in the case that the observed Ti originates from $^{44}$Ti purely. From the above, we conclude that the observed Ti line in the NE jet does not come from $^{44}$Ti.
    
    As the main source contributing to the Ti emissions in the jet structure, we propose a stable Ti isotope, $^{48}$Ti (after decay of $^{48}$Cr). Ti has five stable isotopes, of which $^{48}$Ti is the most abundant, comprising around 90\% of Ti in CC SN models \citep{2021Natur.592..537S}. This stable Ti is produced by both incomplete Si burning and $\alpha$-rich freezeout \citep{Thielemann1996,2001ApJ...555..880N,2021Natur.592..537S}. Therefore, the synthesized amount of stable Ti does not purely reflect the properties of $\alpha$-rich freezeout, which is different from the case of $^{44}$Ti. In the case of $\alpha$-rich freezeout, the $^{44}$Ti and stable Ti are synthesized with almost the same amount. This implies that Ti produced by only $\alpha$-rich freezeout would be difficult to explain our jet observation, since a huge amount of $^{44}$Ti in the jet (larger than the total $^{44}$Ti mass in the whole remnant) is needed in this case. In addition, the elemental composition in the jet, indicating Si-rich ejecta \citep{Hwang2004,2012ApJ...746..130H,2004NewAR..48...61V,Laming2006}, is different from that produced by $\alpha$-rich freezeout. Thus, we propose a possibility that the jet structure was produced by the incomplete Si burning regime.
    \par 

    \subsection{Constraint of the Burning Regime}
    Comparing the observed mass fractions with those of theoretical calculations would be a straightforward approach to understand the nucleosynthetic origin of the jet. In Table \ref{tab:jet_abun} and Fig.~\ref{el_ca}, we summarize the observed mass ratios in the jet and their comparison with theoretical calculations, supporting the incomplete-Si-burning origin for the jet composition. In Fig.~\ref{el_ca} (a), we find global consistency from Si to Ni between the observations and the models in \citet{Thielemann1996}. 
    The initial mass of the Cas A's progenitor is inferred to be in the range of 15--25 $M_{\solar}$ \citep{Chevalier2003,Young2006}. Thus, we choose the SN models in this range for the comparison. Here, we show only the main products at the incomplete Si burning layer ($^{28}$Si, $^{32}$S, $^{36}$Ar, $^{40}$Ca, $^{48}$Cr, $^{52}$Fe, $^{54+56}$Fe and $^{58}$Ni). The elements synthesized efficiently in this layer such as Ar, Ca, Ti and Cr fit better with the models for the 15--20 $M_{\solar}$ progenitors. On the other hand, the models underestimate the mass ratios relative to Ca in some elements (Si, S, Fe and Ni) that are efficiently produced at other layers. This implies that contributions from other burning layers (e.g. explosive O burning, $\alpha$-rich freezeout) to the jet composition might be small. For further discussion, we compare the observed Cr/Fe and Ar/Fe mass ratios with those within the Si burning layers ($T_{\rm peak}>$ 4.5 GK; Fig.~\ref{el_ca} b). In the Si burning layers, the mass fractions are characterized by only a few physical parameters (e.g. $T_{\rm peak}$, $\rho_{\rm peak}$, $Y_{\rm e}$), which is useful to determine the burning regime for the jet formation \citep{2008ApJ...680L..33B,2013ApJ...767L..10P,2017ApJ...834..124Y,2020ApJ...893...49S,2021Natur.592..537S}. 
    In particular, the synthesized amount of the intermediate-mass $\alpha$ elements until Fe (such as $^{36}$Ar, $^{40}$Ca, $^{52}$Fe = $^{52}$Cr) are sensitive to the peak temperature during the nuclear burning (Fig.~\ref{el_ca} b). As a result, we find that the observed mass ratios agree well with those in the incomplete Si burning (quasi statistical equilibrium: QSE) layer with the peak temperature of $\approx$ 5 GK. All of these comparisons with theoretical calculations support that most of the ejecta in the NE jet had been synthesized by the incomplete Si burning regime. This picture also support the optical observations well \citep{2013ApJ...772..134M,2016ApJ...818...17F}. 
    
    \subsection{Jet Formation Mechanism}
    
    What is the actual mechanism that formed the jet structures in Cas A? This would be still an open question. As already well discussed in the previous studies \citep{2014Natur.506..339G,2016ApJ...818...17F}, our results also do not support a jet-driven explosion. In the case of such explosions from some strong central activities, iron-rich ejecta (i.e. $\alpha$-rich freezeout product) could be found along the jet structure or further ahead. However, we can see only the Si-rich ejecta that seem to be produced by the QSE layer. Currently, it would be difficult to explain the widely extended Si-rich (QSE) structure based on our understandings of jet-driven explosions. Recently, \cite{2020A&A...642A..67T} have demonstrated that the Si-rich jet-like features as seen in Cas A can be realized by a post-explosion large-scale anisotropy (= a large clump) at the internal Si-rich layer in the ejecta. If such anisotropy can be created bipolarly in the supernova, it may be able to explain the jet structure seen in Cas A. However, we do not have any plausible mechanisms to create such bipolar large-scale anisotropy at the Si-rich layer of supernovae for now. To create such a bipolar feature, it would be easier to assume a MHD jet or a B-field-modified wind from its proto-neutron star (\cite{2005ASPC..332..350B}). On the other hand, this mechanisms may also require a high-temperature jet and further studies would be needed to understand it. As discussed above, currently we do not have any plausible candidates for explaining all the properties of the Cas A jet. Further theoretical researches would be needed for explaining such a ``weak'' jet that has a low peak temperature ($T_{\rm peak} \lesssim$ 5 GK) during its formation. The estimation of the peak temperature, combined with the jet energy of $\sim$10$^{48}$ up to 10$^{50}$ erg \citep{Laming2006,2008ApJ...686..399S,2016ApJ...818...17F}, will provide a strong constraint on the formation process of the jet structure. To test jet mechanisms comparing with the observational properties will be helpful to understand the origin of this peculiar feature in this remnant.
    
    The $r$-process nucleosynthesis predicted in jet-like explosions of magnetorotational core-collapse supernovae \citep{2015ApJ...810..109N,2017ApJ...836L..21N} would be also useful to discuss the elemental composition in the NE jet. In particular, the observed Ti- and Cr-rich ejecta may be explained by $r$-process nucleosynthesis in an extreme high-entropy and neutron-rich environment. However, there appears to be a challenge in the high-temperature jet mechanism to leave a certain amount of light elements (such as Si, S) as observed. 

\section{Summary}
The origin of the jet structure of Cas A, the most obvious bipolar structure of all the observed remnants, has been debated for years. In this research, we have discovered the emissions from highly ionized Ti in the NE jet of Cas A with {\it Chandra}, which provides us unique information to elucidate its origin. We have estimated the Ti mass is (0.44--2.4)$\times 10^{-5}$ $M_{\solar}$ in the jet region. The estimated Ti mass is too large to be explained by only radioactive $^{44}$Ti, therefore we have concluded that the emissions mainly originate from stable Ti isotopes possibly $^{48}$Ti. While the stable Ti is produced at both the incomplete Si burning and $\alpha$-rich freezeout regimes, the observed elemental composition in the jet region supports the incomplete Si burning origin very well. Thus, it does not support extremely hot jets that could be launched from around its proto-neutron star. We have estimated that the peak temperature during the nuclear burning for the jet formation should be $T_{\rm peak} \lesssim$ 5 GK at most, which would tightly limit the energy of the jet. Currently, we do not have a plausible model to explain all the observational properties of the jet structures. We hope that future theoretical approaches will shed light on this origin.

\section*{Acknowledgment}
We acknowledge helpful discussions with and input from Nozomu Tominaga, Ryo Sawada, Takashi Yoshida, Hideyuki Umeda, Shigehiro Nagataki and Masaomi Ono. This work was supported by KAKENHI Grant Numbers JP18H03722, JP18H05463, JP19K14739, JP20H01941 and JP20K20527.

\bibliography{apssamp}

\end{document}